\documentclass[a4paper,11pt]{article}
\pdfoutput=1
\usepackage{jcappub} 
\usepackage[strings]{underscore}

\usepackage{enumitem}
\usepackage{subcaption}

\title{About Detecting Very Low Mass Black Holes in LAr Detectors}

\author[a]{Ionel Lazanu,}
\author[b,1]{Sorina Lazanu,\note{Corresponding author}}
\author[a]{Mihaela P\^{a}rvu}

\affiliation[a]{University of Bucharest, Faculty of Physics, POBox MG-11 Bucharest-Magurele, Romania}
\affiliation[b]{National Institute of Materials Physics, POBox MG-7 Bucharest-Magurele, Romania}

\emailAdd{ionel.lazanu@g.unibuc.ro}
\emailAdd{lazanu@infim.ro}
\emailAdd{mihaela.parvu@unibuc.ro}

\abstract{The nature of dark matter is still an open problem. The simplest assumption is that gravity is the only force certainly coupled to dark matter and thus the micro black holes could be a viable candidate. We investigated the possibility of direct detection of charged micro black holes with masses around and upward the Planck scale (10$^{-5}$ g), ensuring a classical gravitational treatment of these objects in the next generation huge LAr detectors. We show that the signals (ionization and scintillation) produced in LAr enable the discrimination between micro black holes and other particles. It is expected that the trajectories of these micro black holes will appear as crossing the whole active medium, in any direction, producing uniform ionization and scintillation on the whole path.}

\begin{document}
\maketitle
\flushbottom

\section{Introduction}
\label{sec:intro}
 
There is a general consensus that the observable Universe contains 72\% dark energy, 24\% dark matter and only 4\% ordinary matter. Following the arguments of Paul Frampton \cite{Frampton:2020kdj}, the only interaction which we know for certain to be experienced by dark matter is gravity and the simplest assumption is that gravity is the only force coupled to dark matter. There is no strong argument that the weak interaction must be automatically considered in dark matter interactions. The possible candidates for dark matter constituents are axions, WIMPs, baryonic MACHOs, neutrinos, axinos, gravitinos, \textit{etc}. There is no compelling theory explaining non–radiating ``micro'' black holes -- $\mu$BH; such black holes could be part of the dark matter in the Universe. Any possibility to observe or to determine observational limits on the number of $\mu$BHs (independent of the assumption that they radiate) is very useful.

An important issue is to show that black holes do not radiate in some conditions, as an argument to explain that these relic objects can survive from early Universe. We would need to detect the black holes or to have strong indirect evidence of their existence, as well as to show that they do not radiate. Currently we are far from this stage.

In a classical paper, Hawking \cite{Hawking:1971ei} suggested that unidentified tracks in the photographs taken in old bubble chamber detectors could be explained as signals of gravitationally ``collapsed objects'' ($\mu$BH). The mechanism of black hole formation is well known. As a result of fluctuations in the early Universe, a large number of gravitationally collapsed objects can be formed with characteristics determined by gravity and quantum behaviour. For masses above the Planck mass limit of $10^{-5}$ g quantum behaviour is prevented.

The small black holes are expected to be unstable due to Hawking radiation, but the evaporation is not well-understood at masses of the order of the Planck scale. Helfer \cite{Helfer:2003va} has shown that none of the derivations that have been given for the prediction of radiation from black holes is convincing. They argued that all involve, at some point, speculations on the physics at scales which are not merely orders of magnitude beyond any investigated experimentally ($\sim 10^3$ GeV), but at scales increasing beyond the Planck scale ($\sim 10^{19}$ GeV), where essentially quantum-gravitational effects are expected to be dominant and not all the derivations that have been put forward are mutually consistent. 

Given the profound nature of the issues addressed, some disagreement and controversy exist over exactly what has been achieved. Balbinot \cite{Balbinot:1986fx} demonstrated that when a black hole becomes more and more charged, the Hawking radiation decreases and in the limit of maximum charge containment there is no radiation. Certain inflationary models naturally assume the formation of a large number of small black holes \cite{Chen:2004ft} and the GUP may indeed prevent total evaporation of small black holes by dynamics and not by symmetry, just like the prevention of hydrogen atom from collapse by the standard uncertainty principle \cite{Adler:2001vs}. Chavda and Chavda \cite{Chavda:2002cj} introduced a different idea: gravitationally bound black holes will not emit Hawking radiation. They examine the range $10^{-24}$ kg $\le M_{BH}\le 10^{-12}$ kg where quantum aspects must be considered. These limits of masses are controversial regarding the stability of the black holes, see for example \cite{Rabinowitz:2005ii}.

In this paper we present arguments in favour of the possibility of direct detection of charged $\mu$BHs in LAr detectors, based on their electromagnetic interactions in the material of the target. The case of LXe target is discussed for comparison.

\section{Detection mechanisms}
\label{Detection}
\subsection{Neutral or charged micro black holes?}

Hawking \cite{Hawking:1971ei} argues that black holes can carry an electric charge up to $|Z| \le 30$ in electron units. Zaumen \cite{Zaumen1974} and Christodoulou and Ruffini \cite{Christodoulou1971} used different arguments to predict the electric charge and a limit for it for black holes. Starting from the existence of Hawking radiation \cite{Hawking1975}, Lehmann and co-workers \cite{Lehmann:2019zgt} discuss a possible scenario of accumulation of electric charge in black holes. In this scenario, the black hole evaporates, it emits charged particles of both signs, and it does so stochastically. Thus, during the evaporation process, non-zero electric charges are generated. If evaporation is cut off sharply at some mass scale in the order of $M_{Pl}$, the black hole might be frozen with leftover electric charge of a random sign.
Depending on the electric charge, these primordial $\mu$BH can capture electrons or protons to form neutral ``atoms'', but also starting from this state the stripping processes can be considered. According to Hawking [2], the ``process of neutralisation would continue until the temperature falls to the point where the wavelength of the particles was greater than the radius of the hole''.

In the case of any projectile with mass and charge of a certain sign, surrounded by particles of the opposite sign (electrons or protons), determining the average equilibrium charge during penetration through matter is a difficult problem. The mean equilibrium charge can be estimated on the basis of either Bohr’s velocity criterion or of Lamb’s energy criterion \cite{Sigmund2014}. A dynamical equilibrium is eventually approached around a charge state for which the number of capture events equals the number of loss ones per time or particle range. 

If the charge of the projectile is higher than the equilibrium charge, capture typically dominates over loss, and vice versa in the opposite case, charge exchange being a stochastic process.  For an individual trajectory, charge fractions can take any value that is compatible with conservation laws. In the present discussion we use the velocity criterion. 
Consider a $\mu$BH moving with a velocity $v$ and carrying a number of electrons with orbital speeds $v_e$. The electrons of the $\mu$BH could be divided according to Bohr \cite{Bohr1940} into two groups: those with orbital speed greater and smaller than $v$, respectively. Those with $v_e < v$ are relatively weakly bound and could be stripped, while those with orbital velocities significantly greater than $v$ would respond adiabatically to the disturbance, having this way a small chance to be stripped off. By extrapolating the original idea of Bohr \cite{Bohr1948}, Pierce and Blann \cite{Pierce1968} found for the equilibrium charge the formula

\begin{equation}
q \cong Z_1 \left(1-e^{-v/Z_1^{2/3} v_{Bohr}}\right).
\end{equation}
in the hypothesis of the hydrogenic model for the projectile, $v_{Bohr}$ being Bohr’s velocity.

Up to this point, the charge state of the moving projectile has been considered as independent of the medium. The effects of the medium must be taken into account, and they are associated with a new concept, of effective electric charge, a quantity depending on the charge and velocity of the projectile and on the electric charge of the target atoms ($q \to Z_{eff}$). The concrete procedure to define the effective charge of the projectile is through the square of the ratio between the stopping power of the projectile ($\mu$BH or ion) and the stopping power of hydrogen in the medium, while the effective charge of the hydrogen depends also on the velocity of the projectile \cite{Ziegler}. 
In all calculations associated with the energy losses of the $\mu$BH as well as of recoil nuclei in Ar and Xe respectively, this quantity was used. 

\begin{figure}[h]
\centering
\includegraphics[scale = 1]{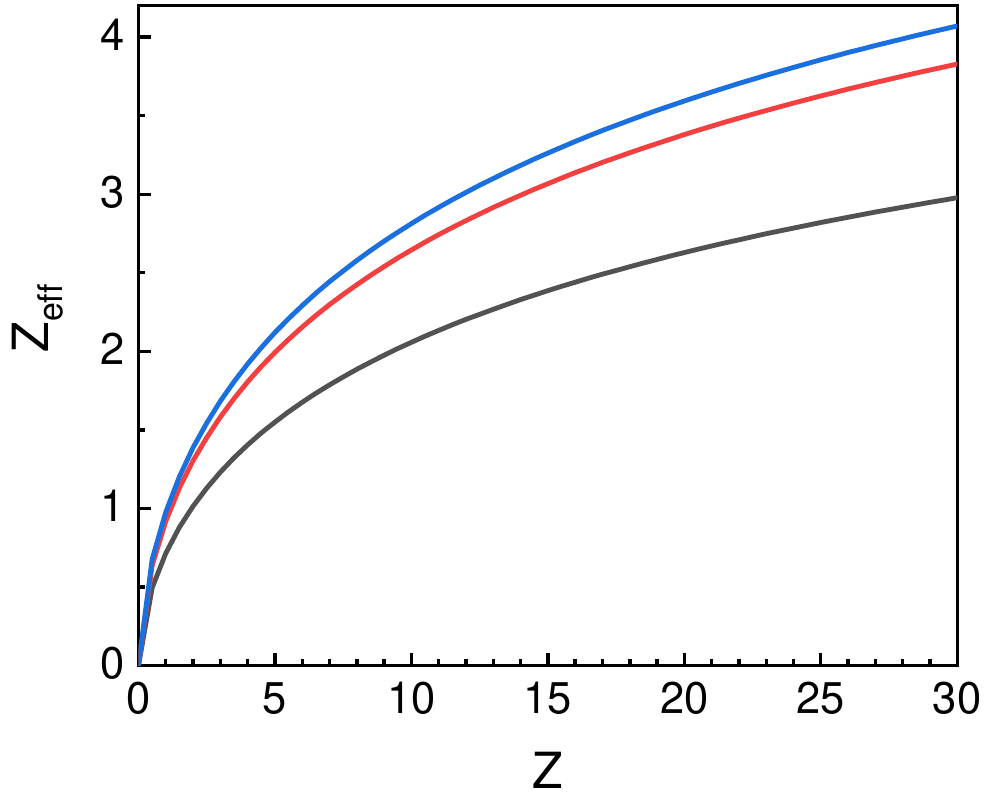}
\centering
\caption{$Z_{eff}$ vs. $Z$ for a $\mu$BH of $10^{-5}$ g mass and for the velocities, from bottom to top, of 250, 500 and 750 km/s in LAr.}
\label{Fig1}
\end{figure}

The dependence of the effective charge of the $\mu$BH, $Z_{eff}$, on its charge $Z$ is represented in Figure \ref{Fig1} for the particular case of a $\mu$BH of $10^{-5}$ g mass and for the velocities of 250, 500 and 750 km/s in LAr. A tendency toward saturation is found at increasing Z. 

\subsection{Rates}

Recent estimates from global methods for dark matter densities lie in the range $0.2 - 0.6$ GeV/cm$^3$. A major source of uncertainty is the contribution of baryons (stars, gas, and stellar remnants) to the local dynamical mass. New studies of the local DM density from Gaia satellite data yield $0.4 - 1.5$ GeV/cm$^3$, depending on the type of stars used in the study. An updated halo model predicts values of $0.55 \pm 0.17$ GeV/cm$^3$, where the 30\% error accounts for the systematics \cite{PDG}. Other observational quantities of interest are the local circular speed \textit{v$_c$} and the escape velocity \textit{v$_{esc}$}. The values obtained are $v_c = 218 - 246$ km/s and $v_{esc}=533^{+54}_{-41}$ km/s \cite{PDG}.
We expect these very low mass objects to have velocities in the range $250 - 1000$ km/s because their primary motion is under the effect of the gravity.

In the hypothesis that the dark matter only consists of $\mu$BH objects, the flux can be calculated as

\begin{equation}
\Phi_{\mu BH} \cong \frac{\rho_{DM}}{M_{\mu BH}} v_{DM} \approx 1.6\times[10^{-16} - 10^{-35}]\, \mathrm{part/(m^2 s)} .
\end{equation}
where $\rho_{DM}=0.35$ GeV/cm$^3$, $v_{DM}=250$ km/s and $M_{\mu BH}={10}^{-5} - {10}^{14}$ g.

The interaction of the neutral black hole atoms with ordinary matter via the gravitational dynamical friction effect is extremely weak, as it was first shown in \cite{Chen:2004ft}. This is due to the extremely small cross-section, of the order of $ \sim \pi r_{g}^{2} \times (c/v)^2 \sim 3 \times 10^{-60}$  m$^2 $ \cite{Dokuchaev:2014vda} .

In a simple approximation, the total rate \textit{R} is 
\begin{equation}
R  = \mathrm{flux} \times \mathrm{number~ of~ target~ particles} \times \mathrm{cross~ section}.
\end{equation}

In accordance with Hawking \cite{Hawking:1971ei}, for elastic collisions of neutral black hole atoms, the cross section is of the order of 10$^{-20}$ m$^2$ for an electronic atom and 10$^{-27}$ m$^2$ for a protonic one. The number of target particles ($N_t$) is calculated as $N_t=N_A\frac{M_t}{A}$, where $N_A$ is Avogadro’s number and (\textit{A}, \textit{M$_t$}) are the mass number and mass of the target (detector). Thus the rate of events is

\begin{equation}
R\left[\mathrm{part./year}\right]=7.6\times M_t\left[\mathrm{kg}\right]\times\ \begin{cases}
{10}^{-48}  \mathrm{\; gravitational\; interaction}\\
{10}^{-8} \mathrm{\; elastic;\;  for\; ``electronic\; atom"}\\
{10}^{-15} \mathrm{\; elastic;\;  for\; ``protonic\; atom"}
        \end{cases}
\end{equation}
where the last column is the multiplicative factor considering the gravitational interaction, elastic scattering for the ``electronic atom'' and the ``protonic atom'' respectively. Starting from the argument of Hawking \cite{Hawking:1971ei} previously exposed, we suppose that as the $\mu$BH passes through the detector it does not change its electrical charge through loss or capture of electrons (or protons for $\mu$BHs with negative electric charge).

Liquefied noble gases have a unique combination of properties: a) scintillation; b) possibility that electrons released in the ionization process remain free to drift across long distances; c) possibility of extracting electrons to the gas phase, where the ionization signal can be amplified through secondary scintillation or avalanche mechanisms in double phase detectors; d) can be produced in large quantities. Liquid noble gases are dense materials, adequate to obtain homogeneous targets and also detectors that do not attach electrons, are inert (not flammable) and very good dielectrics. From them, LAr and LXe are mostly used as materials for detectors. They can be easily obtained and purified commercially and in principle large detector masses are feasible (at modest costs compared to semiconductors). The liquid phase is obtained at 170 K in LXe and 87 K in LAr respectively. In this paper, we consider the LAr option, and we present in Table \ref{table:1} a compilation of present experiments and future projects of neutrino physics able to detect $\mu$BHs. Currently, the state of the art in LXe detectors is represented by the experiment XENON 1T, while XENON nT (8 tons) is expected in 2023.

\begin{table}
\begin{center}
\begin{tabular}{  |p{3cm}|p{2cm}|p{4cm}|p{5cm}| } 
\hline
\textbf{Detector}& \textbf{Reference} & \textbf{Technology} & \textbf{Active Mass of Argon} \\ 
\hline
DarkSide & \cite{darkside} & Dual-Phase & 33 kg \\
\hline
ArDM & \cite{ardm} & Dual-Phase & 850 kg \\
\hline
MiniCLEAN & \cite{miniclean} & Single-Phase & 500 kg \\
\hline
DEAP & \cite{deap} & Single-Phase & 3600 kg \\
\hline
ArgoNeuT & \cite{argoneut} & Single-Phase & 0.02 ton \\
\hline
MicroBooNE & \cite{microboone} & Single-Phase & 170 ton \\
\hline
ICARUS & \cite{icarus} & Single-Phase & 600 ton \\
\hline
ProtoDUNE-SP & \cite{protodunesp} & Single-Phase & 770 ton \\
\hline
ProtoDUNE-DP & \cite{protodunedp} & Dual-Phase & 760 ton \\
\hline
DUNE & \cite{Abi:2020wmh} & Single+Dual-Phase & up to 4 $\times$ 10 kton \\
\hline
\end{tabular}
\caption{Current Ar experiments and future projects.}
\label{table:1}
\end{center}
\end{table}

For many of the experiments listed in Table \ref{table:1}, the masses of the targets are insufficient to achieve a significant number of events. In these very rough estimations, around 3 events/ year are expected for the DUNE detector with all 4 modules.

\subsection{Gravitational, electronic and nuclear energy loss}

When a projectile interacts with Ar, it loses energy by gravitational interactions, and/or transfers energy to atomic electrons producing ionizations and excitations, and/or interacts elastically directly with nuclei as Coulomb processes. 

The gravitational effects in the transmission of $\mu$BH through matter can be estimated in the classical impulse approximation using the analogy with the Bohr formalism for ions. The $\mu$BH energy loss per unit length is
\begin{equation}
\label{eq:dEdx}
\frac{dE}{dx}=\frac{4\pi G^2M^2\rho}{v^2} \mathrm{ln}\left(\frac{b_{max}}{b_{min}}\right) \sim \frac{4\pi G^2M^2\rho}{v^2}\times\left(10\div30\right). 
\end{equation}
where \textit{G} is Newton’s constant, $\rho= m \times n$ is the mass density of the target particles with \textit{n} the number density of the target particles. The term ln$\left(\frac{b_{max}}{b_{min}}\right)$ changes weakly for a very large domain of impact parameters due to the logarithmic dependence. A value between 10 and 30 was suggested by Rabinowitz \cite{Rabinowitz:2005ii}. 

At very low energies, the interaction of the $\mu$BH with nuclei gets dominant compared to the contribution of the interaction with electrons. In a second step, recoil nuclei produced in the primary interaction also interact with electrons and other nuclei up to their stopping in LAr.

Electronic and nuclear stopping powers are calculated using the procedure used by Lazanu and Lazanu in previous papers \cite{Lazanu:2011zn,Lazanu:2011qi}. The energy transferred in electronic processes is used for the ionization and for the excitation of the electrons. 

In fact, one can consider the LAr as consisting of two subsystems: electronic and nuclear. The energy lost by the projectile is partitioned between these subsystems, which have different energy distributions, equivalent to two different temperatures. For a very short period of time a transfer of energy between subsystems must be considered simultaneously with the diffusion of heat in space. 

In this case, the primary interaction in LAr, seen as a binary process, takes place under the condition  $M \gg m_i$, with $i= \mathrm{Ar}, e$; in both cases the maximum transferred energy is $E_{K,max} \approx 2m_ic^2\beta^2\gamma^2$.

For velocities of $\mu$BHs of $250-1000$ km/s, the kinetic energy transferred in an interaction to electrons is of the order $1-10$ eV and $51.7-827.2$ keV to Ar nuclei respectively. The ionization potential in the liquid phase is 13.4 eV. The average energy required to produce an electron-ion pair in LAr is $W_i=23.6 \pm 0.3$ eV, a value that is smaller than that obtained in gaseous Ar $W_g= 26.4 \mathrm{\; eV}$. In these circumstances $\mu$BHs have a small contribution to excitation and ionization processes, while nuclear processes are dominant. 

\subsection{The electronic stopping power}
In this low energy range of the $\mu$BHs of interest here, the electronic stopping of the charged incoming particle in the target is treated within the local density approximation; the ion-target interaction is treated as that of a particle with a density averaged free electron gas. 
The basic assumptions of this approximation are: a) the electron density in the target varies slowly with position; b) the available electron energy levels and transition strengths are described by those in a free electron gas and there are no significant band-gap effects; c) the charge of the ion can be reduced to a scalar quantity called the ion's effective charge.

At low energies, Lindhard’s formula \cite{Lindhard} is used, where the energy loss is proportional to the velocity. 

The electronic (masic) stopping power is
\begin{equation}
S_{el}\left(E\right)=\frac{1}{\rho}\times\left(\frac{dE}{dx}\right)_{el}=2.307\times{10}^9\times\frac{Z^{7/6}Z_{Ar}}{\left(Z^{2/3}+Z_{Ar}^{2/3}\right)^{3/2}}\times\frac{1}{A_{Ar}}\times\left(\frac{E}{A}\right)^{1/2} 
\label{Sel}
\end{equation}
in units of $\mathrm{eV} \times \mathrm{cm}^2/\mathrm{g}$. In this equation \textit{E} is given in keV, \textit{A} in atomic mass units, and \textit{Z} is the effective charge of the $\mu$BH.

At high energies, if the velocity of the projectile is $v>v_{Bohr}$ of electrons in the atom, the electronic stopping power is calculated using the  Bethe-Bloch formula.

For charged $\mu$BHs and for ions moving in matter, in the calculus of the electronic stopping their effective charge must be considered and this value varies with the velocity, as specified previously.

In contrast to the charged $\mu$BHs that, having electric charge, interact in matter by Coulomb interactions, neutral particles such as WIMPs do not interact directly with electrons; these exotic particles can interact only as a singular process with nuclei, and ionizations and excitations are produced by the recoil nuclei. Neutral $\mu$BHs behave similarly to WIMPs.

\subsection{The nuclear stopping power}
The nuclear stopping power is calculated using the Ziegler, Biesack, Litmark formalism \cite{Ziegler2}. The reduced energy $\varepsilon$ is defined as

\begin{equation}
\varepsilon=\frac{32.53\ A_{Ar}}{A+A_{Ar}}\times\frac{E}{ZZ_{Ar}\left(Z^{2/3}+Z_{Ar}^{2/3}\right)^{1/2}} .
\label{eps}
\end{equation}
where the energy \textit{E} is expressed in keV. Correspondingly, the reduced nuclear stopping cross section is
\begin{equation}
S_n\left(\varepsilon\right)=\frac{1}{2}\frac{\ln\left(1+\varepsilon\right)}{\varepsilon+0.10718\ \varepsilon^{0.37544}} .
\end{equation}

With these definitions, the nuclear masic stopping power is
\begin{equation}
S_{nucl}\left(E\right)=\frac{1}{\rho}\times\left(\frac{dE}{dx}\right)_{nucl}=5.097\times{10}^9\frac{ZZ_{Ar}}{\left(A+A_{Ar}\right)\left(Z^{2/3}+Z_{Ar}^{2/3}\right)^{1/2}}\times\frac{A}{A_{Ar}}\times S_n\left(\varepsilon\right)
\label{Sn}
\end{equation}
in units of $\mathrm{eV} \times  \mathrm{cm}^2/\mathrm{g}$.
In the first step, when the $\mu$BH interacts directly in the detector, the charge and mass numbers are $Z\rightarrow Z_{\mu BH}$ and $A\rightarrow A_{\mu BH}$; in all the other cases, the projectile is the Ar self-recoil so $Z\rightarrow Z_{Ar}$ and $A\rightarrow A_{Ar}$.

\section{Production of scintillation light in liquid argon}
Following the model of Doke and co-workers \cite{Doke:2002oab}, there are two distinct possible ways for scintillations in Ar
\begin{enumerate}[label=(\roman*)]
\item $R^\ast:\\
R^\ast+R+R\rightarrow R_2^\ast+R\\
R_2^\ast\rightarrow2R+h\nu$
\item	$R^+:\\
R^++R\rightarrow R_2^+ \\
R_2^++e^-\rightarrow R^{\ast\ast}+R \\
R^{\ast\ast}\rightarrow R^\ast+heat \\
R^\ast+R+R\rightarrow R_2^\ast+R \\ 
R_2^\ast\rightarrow2R+h\nu$
\end{enumerate}

In both processes, the excited dimer at the lowest excited level should be de-excited to the dissociative ground state by emitting a single UV photon, because the energy gap between the lowest excitation level and the ground level is so large that there exists no decay channel such as nonradiative transitions. Although this is not yet fully confirmed by experiments, it is generally assumed that each excited dimer emits a single photon.

After the interaction of the incident particle, singlet and triplet dimers will be produced, and the scintillation is a product of the two radiative decays following the excitation process. Liquid and gaseous Ar are transparent to their own scintillation lights. In the case of double phase technology, ionizations and scintillations are produced in both phases. In accordance with \cite{Hitachi:1983zz}, the singlet decays quickly, being responsible for most of the prompt light seen in the scintillation spectrum, whereas the triplet decays with a longer lifetime. The time constants of the singlet and triplet decays have been measured in all phases and their lifetimes are $7.0 \pm 1.0$ ns and $1600 \pm 100$ ns respectively. The existence of different impurities in argon put in evidence the differences in the emission spectra of liquid and gaseous phases \cite{Buzulutskov:2017wau,Heindl:2010zz}. In LAr, the spectrum is dominated by an emission feature of 126.8 nm or 9.78 keV equivalent value analogues to the 2nd excimer continuum in the gas phase \cite{Parvu:2017xde}, confirming the previous results of Doke \cite{Doke:2002oab}. Weak-emission features in the wavelength range from 145 to 300 nm were observed. The structure at 155 nm in the gas phase has only a very weak analogue in the liquid phase. The structure at longer wavelengths up to 320 nm is addressed as the 3rd continuum emission in the gas phase.

\begin{figure}[h]
\centering
\includegraphics[scale = 0.45]{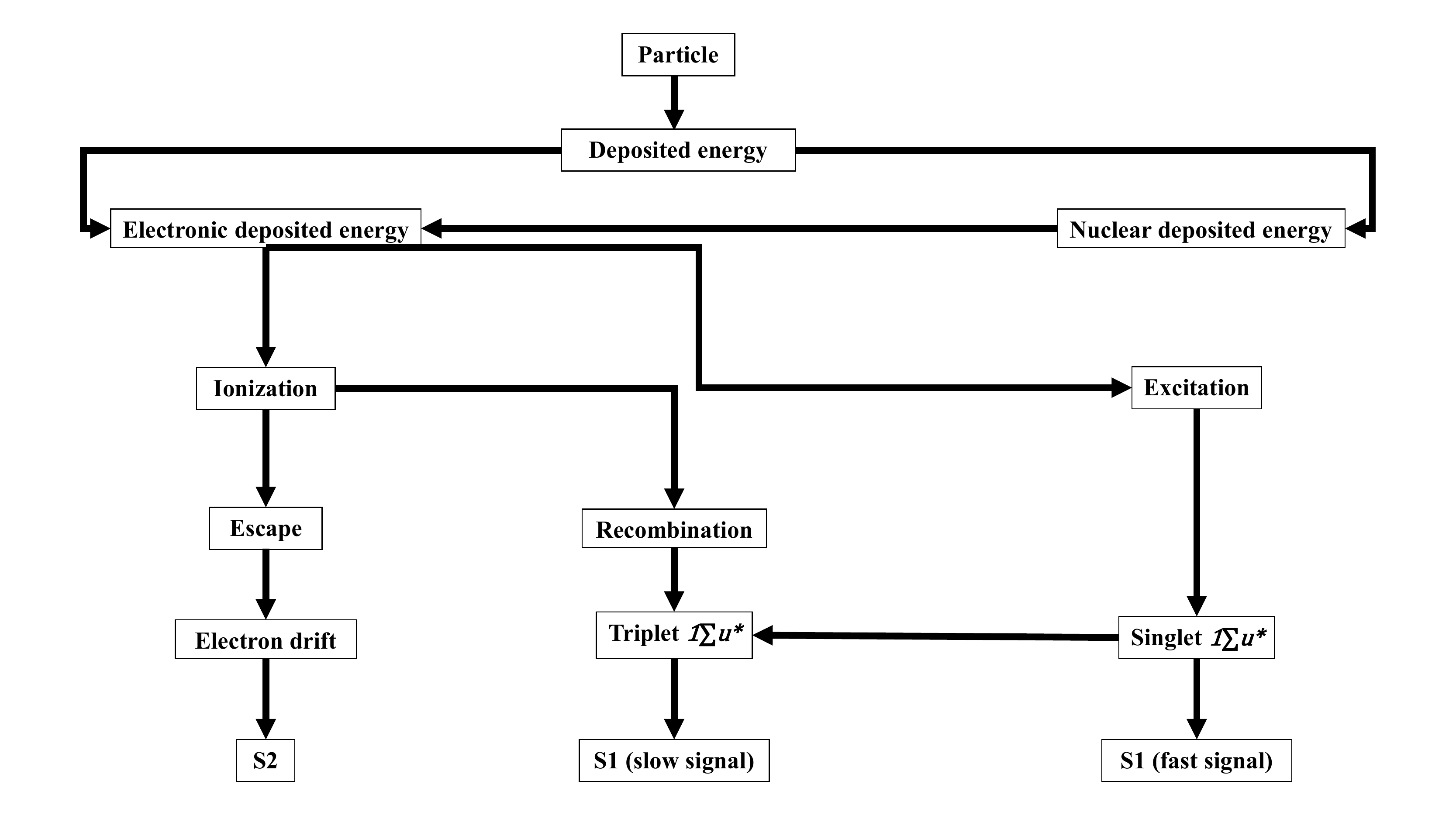}
\centering
\caption{The liquid argon response to interaction of incident charged particle.}
\label{Fig2}
\end{figure}

Inspired by Refs. \cite{Suzuki} and \cite{Kimura:2016nsk}, the response of LAr to electronic excitation is presented in Figure \ref{Fig2}. 

Recently, in different and successive experiments, the ionization yields of nuclear recoils in  LAr have been measured in the low energy range at 6.7 keV \cite{Joshi:2014fna}, $17 - 57$ keV \cite{Cao:2014gns} and at higher energies, at 80 and 233 keV \cite{Bondar:2014nra}. These measurements cover part of the energy range of interest for the present study. Because our analysis is not dedicated to the detection possibilities specific to a particular experiment, we will make only a few general remarks. The S1 and S2 light yields depend on the strength of electric field in the drift interaction region, mainly due to the recombination effect of ionizing electrons. Unfortunately, the basic properties of S1 and S2 of Ar are not well known. Recently, Washimi and co-workers \cite{Washimi:2018kgn, Washimi:2018ugh} focused their studies on the drift-field dependence of S2/S1 properties for electric fields of $0.2 - 3.0$ kV/cm, of interest for two-phase Ar detectors. In figure 2 from reference \cite{Washimi:2018ugh} the drift field dependences of S1 and S2 signals are presented, while the S2/S1 ratio for a two-phase detector is given in reference \cite{Washimi:2018kgn}.

\section{Results and discussions}

The gravitational energy loss for a $\mu$BH of a mass of 10$^{-5}$ g and a velocity of 250 km/s, evaluated using Eq. (\ref{eq:dEdx}) is in the order $0.9\times{10}^{-21}$ keV cm$^2$/g, and in the next discussion the gravitational energy loss will be neglected with respect to the electronic and nuclear energy losses.

\begin{figure}[h]
\includegraphics[scale = 1]{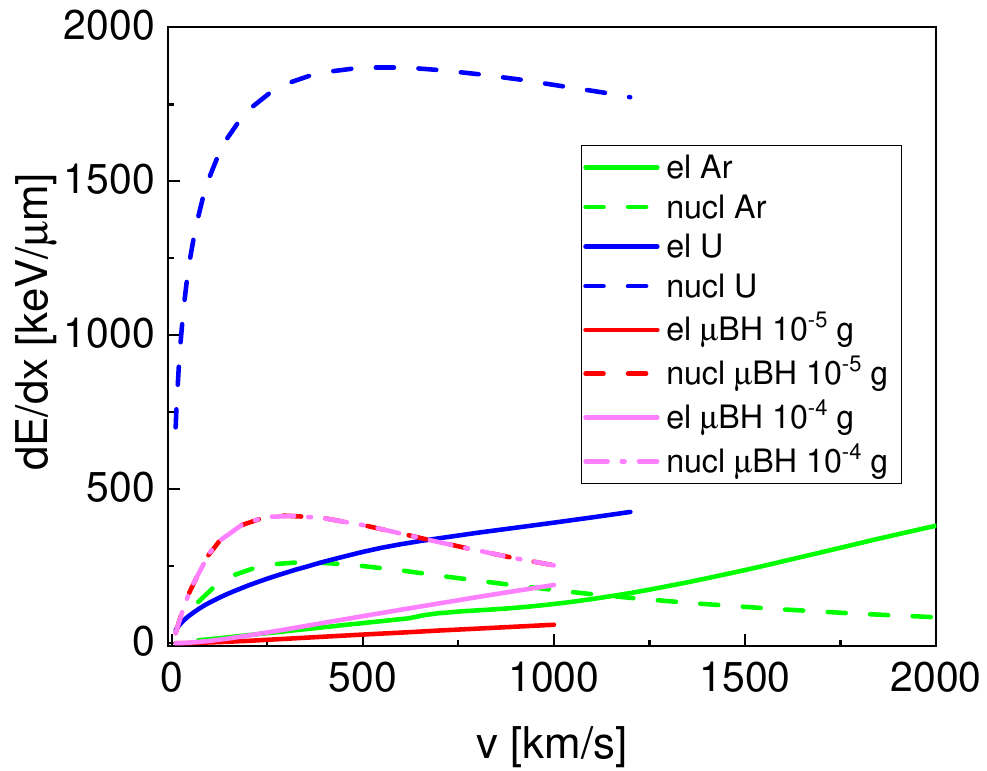}
\centering
\caption{Velocity dependence of the electronic (lines) and nuclear (dashed lines) energy loss for the $\mu$BH of $10^{-5}$ and $10^{-4}$ g, for Ar selfrecoils and U ions in LAr.}
\label{Fig3}
\end{figure}

In Figure \ref{Fig3} we present the velocity dependence of the nuclear and electronic energy loss in LAr for the $\mu$BH and for the Ar recoils produced in the interaction of the $\mu$BH in the detector. Two values for the mass of the $\mu$BH are considered: $10^{-5}$ and $10^{-4}$ g; its charge number is 30.

As we specified earlier, for WIMPs interactions there is no electronic energy loss produced by the direct interaction, but only that due to recoil-induced processes. For comparison with the $\mu$BH, the energy loss of a heavy ion (uranium was considered as example) in LAr is also included in the figure, with the same velocities as the $\mu$BH, allowing us to highlight the differences expected in the signals resulting from the interactions. One can see that the U ion has higher energy loss than the $\mu$BH, both nuclear and electronic. The energy losses for the $\mu$BH were calculated using Eqs. (\ref{Sel}) and (\ref{Sn}), while the values for Ar and U are from SRIM \cite{ZieglerSRIM}. The results show the peculiarities of the energy losses of $\mu$BHs both in electronic and nuclear processes, compared with Ar self-recoils and with U ions. For $\mu$BHs and for U ions in LAr, the nuclear energy loss is greater than the electronic energy loss in the range of velocities investigated. For recoil Ar nuclei, the electronic energy loss is greater than the nuclear energy loss above 1200 km/s. The increase in the mass of the $\mu$BH determines the increase of its electronic energy loss, while the nuclear one is kept unchanged.

A special mention must be done for the variation of the effective charge of the $\mu$BH in LAr. Considering that the $\mu$BH of $10^{-5}$ g and 250 km/s velocity moves through a LAr detector of around 60 m length, as is the case of DUNE detector, the relative variation of its velocity is of only about $10^{-6}\%$. Therefore, it is justified to consider its effective electric charge as constant.

\begin{figure}[h]
\includegraphics[scale = 1]{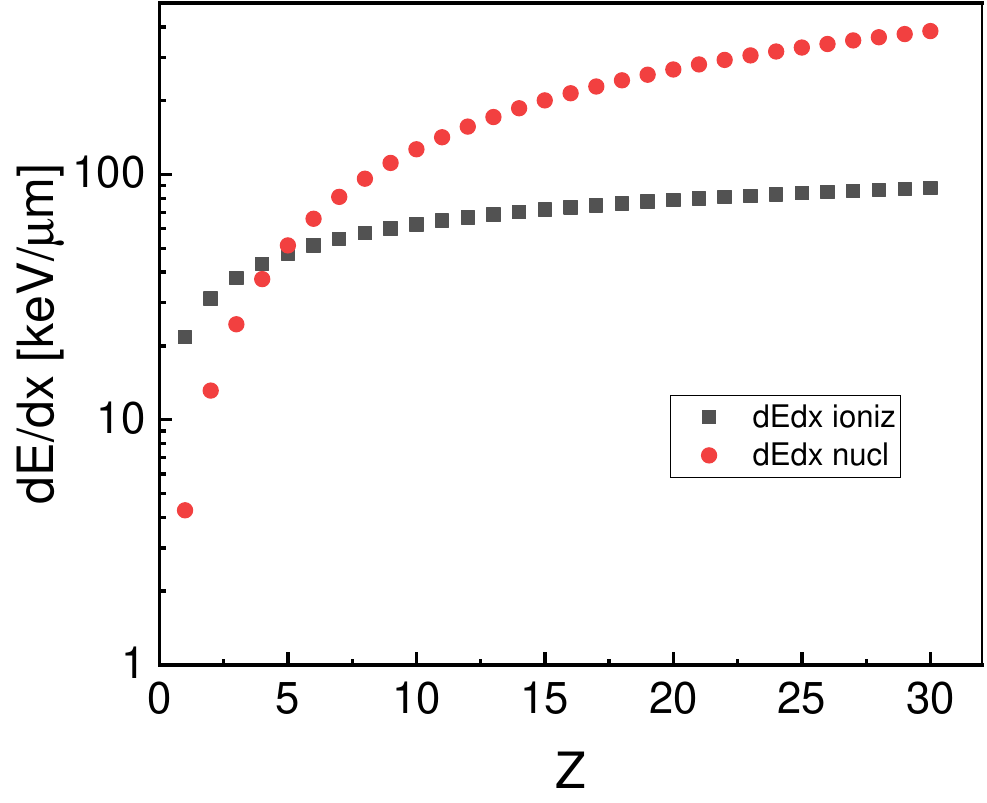}
\centering
\caption{Electronic and nuclear energy losses for the $\mu$BH of $10^{-4}$ g and 500 km/s versus its charge number in LAr.}
\label{Fig4}
\end{figure}

In Figure \ref{Fig4} we present the dependence of the electronic and nuclear energy losses of a $\mu$BH of mass $10^{-4}$ g and velocity 500 km/s in LAr on its charge. Both ionization and nuclear energy loss present a saturation behaviour with the increase of the charge number. 

Transient thermal processes follow the scattering process, when an appreciable amount of energy is released in a small region of a material, in a very short time interval. There are many models in the literature for these processes, one of them being the thermal spike model. In this model, the target is considered to have two subsystems, namely electrons and nuclei, which are coupled through the energy transfer, which could take place in both directions. 

Heat diffusion in both subsystems is described by the classical heat equation, supposing the existence of two sources: one source is given by the energy released through electronic energy loss, the other by nuclear energy loss. The consideration of two sources is required in the regime of comparable electronic and nuclear energy losses. This thermodynamic model is not the most appropriate way to follow, but the only one for which the formalism allows an unambiguous approach. A detailed description of the model, equations and steps to solve the time and space evolution of both subsystems is discussed in reference \cite{Lazanu:2015qpa}. Nuclear energy loss is dominant at the end of range, or at low kinetic energies of the projectile, while the electronic stopping becomes increasingly important at high energies. Both recoils and secondary electrons produce heating effects while spreading their energy.

The thermal spike model, which we use in this work, was developed initially by Koehler and Seitz \cite{KoehlerSeitz} for effects produced by irradiation in crystals and by Seitz \cite{Seitz:1958nva} to explain bubble formation due to charged particles interacting in liquids. Toulemonde \textit{et al.} \cite{Toulemonde} have applied the thermal spike model to water. The electronic and atomic (or molecular) sub-systems in the LAr target have different temperatures, and are coupled through a term proportional to the temperature difference between them. The energy depositions around the trajectories of ions in fluids were modelled as thermal spikes and pressure waves using the equations of fluid dynamics by Apfel \textit{et al.} \cite{Apfel}; in this model, only the energy transferred to the atomic system is considered by the authors, the electronic component being neglected.

The dependencies of the temperatures of electron and molecular subsystems on the distance to the track of the projectile (recoil), \textit{r}, and on the time after its passage, \textit{t}, are solutions of two coupled partial differential equations \cite{Lazanu:2011qi}
\begin{equation}
\begin{matrix}C_e\left(T_e\right)\frac{\partial T_e}{\partial t}=\frac{1}{r}\frac{\partial}{\partial r}\left[rK_e\left(T_e\right)\frac{\partial T_e}{\partial r}\right]-g\left(T_e-T_{mol}\right)+A\left(r,t\right)\\C_{mol}\left(T_{mol}\right)\frac{\partial T_a}{\partial t}=\frac{1}{r}\frac{\partial}{\partial r}\left[rK_{mol}\left(T_{mol}\right)\frac{\partial T_{mol}}{\partial r}\right]-g\left(T_{mol}-T_e\right)+B\left(r,t\right)\\\end{matrix}
\end{equation}
where \textit{T$_{e(mol)}$} , \textit{C$_{e(mol)}$}, and \textit{K$_{e(mol)}$} are respectively the temperatures, the specific heat, and the thermal conductivities of the electronic (index $e$) and molecular subsystems (index \textit{mol}). 

The sources satisfy the conservation laws
\begin{equation}
\begin{matrix}\int_{0}^{\infty}dt\int_{0}^{\infty}{2\pi rA\left(r,t\right)\ dr=\left(\frac{dE}{dx}\right)_{el}}\\\int_{0}^{\infty}dt\int_{0}^{\infty}{2\pi rB\left(r,t\right)\ dr=\left(\frac{dE}{dx}\right)_n}\\\end{matrix}
\end{equation}
with $\left(\frac{dE}{dx}\right)_{el}$ and $\left(\frac{dE}{dx}\right)_n$ the electronic and nuclear linear energy losses, respectively. 
While the thermodynamic parameters for the molecular subsystem of LAr were extensively investigated, the ones corresponding to the electronic subsystem are not known. In the numerical calculations, the values of \textit{C$_{mol}$} and \textit{K$_{mol}$} for LAr were taken from Ref. \cite{Aprile:2008bga}, while for the electronic one the values from Ref. \cite{KoehlerSeitz} were used. For the coupling constant between the sub-systems, \textit{g}, related to the mean free path of electrons, a value of 10$^{13}$ W cm$^{-3}$K$^{-1}$ was taken. 

In LAr, it is generally agreed that the energy deposited into the electronic subsystem is divided between ionization and scintillation, 82.6\% and 17.4\% respectively. Supposing that the applied electric field directs the electrons to the electrode, we assume that only the part not assigned to ionization from the electronic energy loss is available for energy exchange with the molecular subsystem.

\begin{figure}
\centering
\begin{subfigure}{.5\textwidth}
  \centering
  \includegraphics[width=1\linewidth]{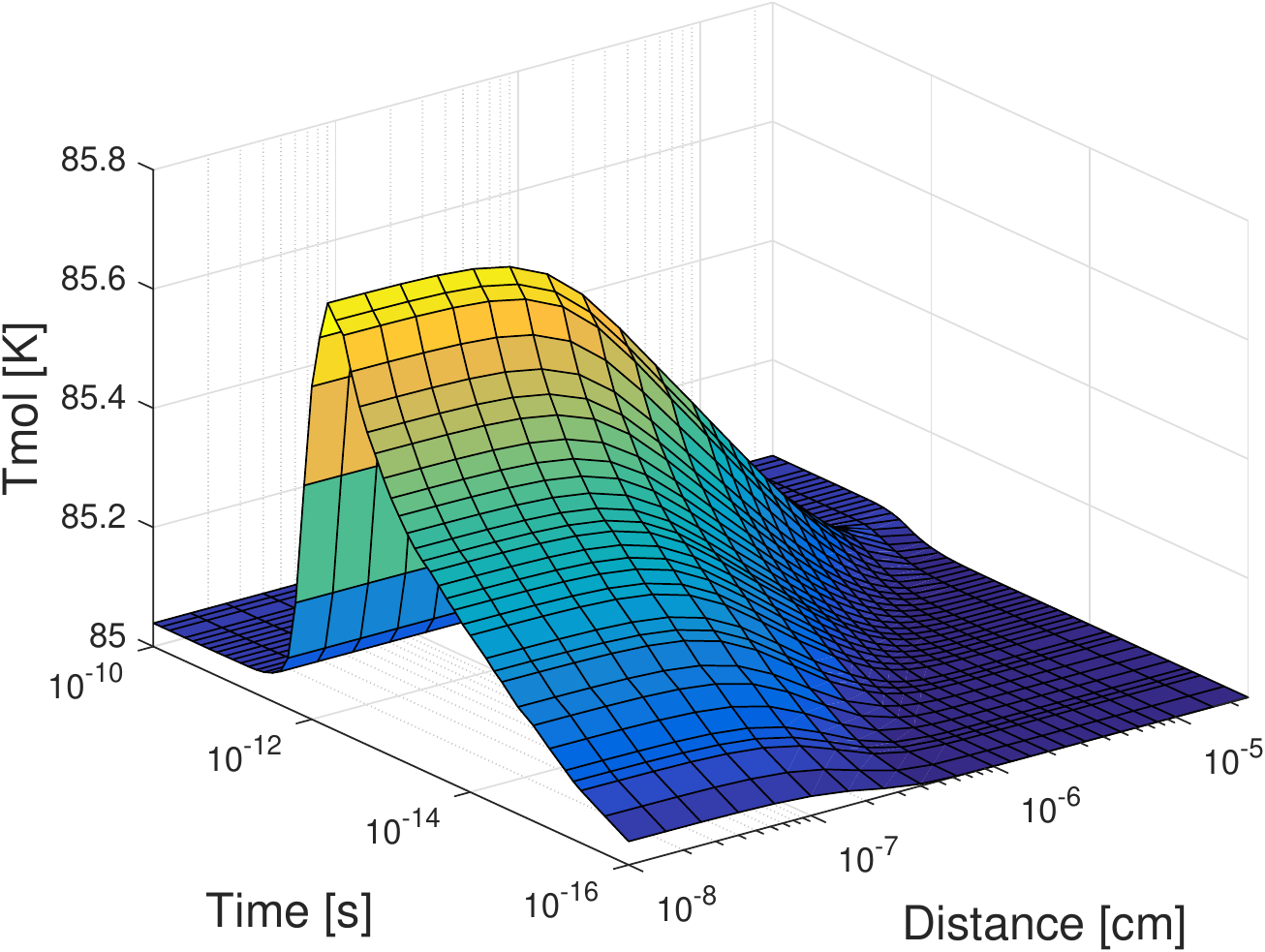}
  \caption{}
  \label{Fig5a}
\end{subfigure}%
\begin{subfigure}{.5\textwidth}
  \centering
  \includegraphics[width=1\linewidth]{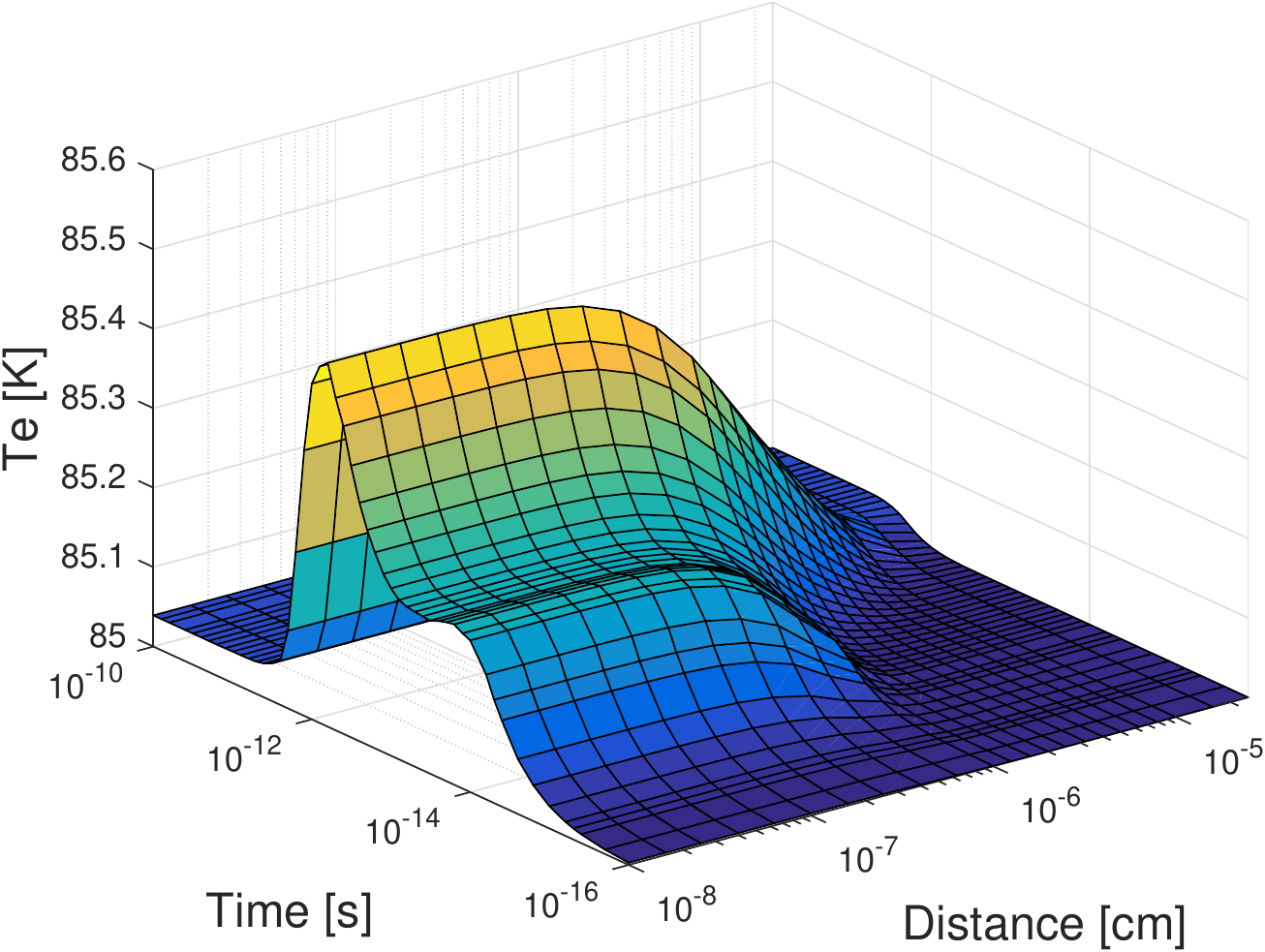}
  \caption{}
  \label{Fig5b}
\end{subfigure}
\caption{Time and space variation of the temperature of the molecular (a) and electronic system (b) in LAr produced by a $\mu$BH of velocity 250 km/s, mass $10^{-5}$ g  and charge number 30.}
\end{figure}

In Fig. \ref{Fig5a} and \ref{Fig5b} the development of the thermal spike in LAr, in space and time is illustrated. It is produced by a $\mu$BH of mass 10$^{-5}$ g and velocity 250 km/s. The electronic and molecular subsystems have different characteristic times, \textit{i.e.} $5 \times 10^{-15}$ and $5 \times 10^{-13}$ s respectively. One can see a small increase of the electronic temperature due to the electronic energy loss, and another, more important increase, due to the transfer from the atomic (molecular) subsystem. 

These results highlight the following aspects of the temporal and spacial evolution of transient phenomena: a) at a time scale of the order of 10$^{-16}$ s and a spacial scale of atomic dimensions, the two subsystems receive energy by transfer from the projectile according to linear energy losses, and keeping into account that only part of the energy transferred to electronic sub-system is available for exchange; b) at a time scale of the order $5 \times 10^{-13}$ s, the atomic subsystem dominantly transfers energy to the electronic one, which uses it to produce additional ionizations and excitations. In the case of $\mu$BH this is the dominant component of ionization and excitation energy; c) in these processes, the temperature increase does not produce a liquid-gas phase transition for an initial temperature of LAr of 85 K.

\begin{figure}
\centering
\begin{subfigure}{.5\textwidth}
  \centering
  \includegraphics[width=1\linewidth]{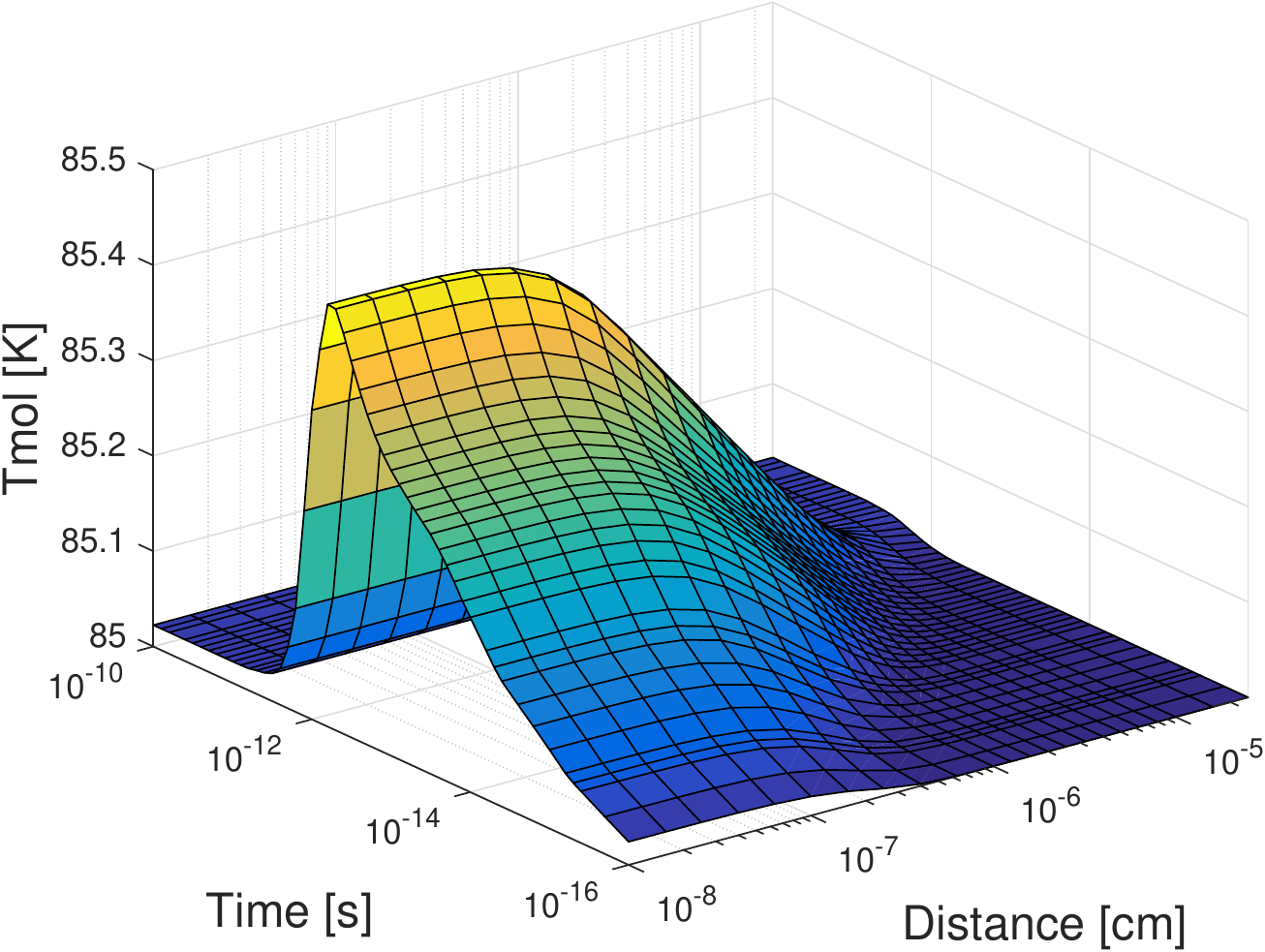}
  \caption{}
  \label{Fig6a}
\end{subfigure}%
\begin{subfigure}{.5\textwidth}
  \centering
  \includegraphics[width=1\linewidth]{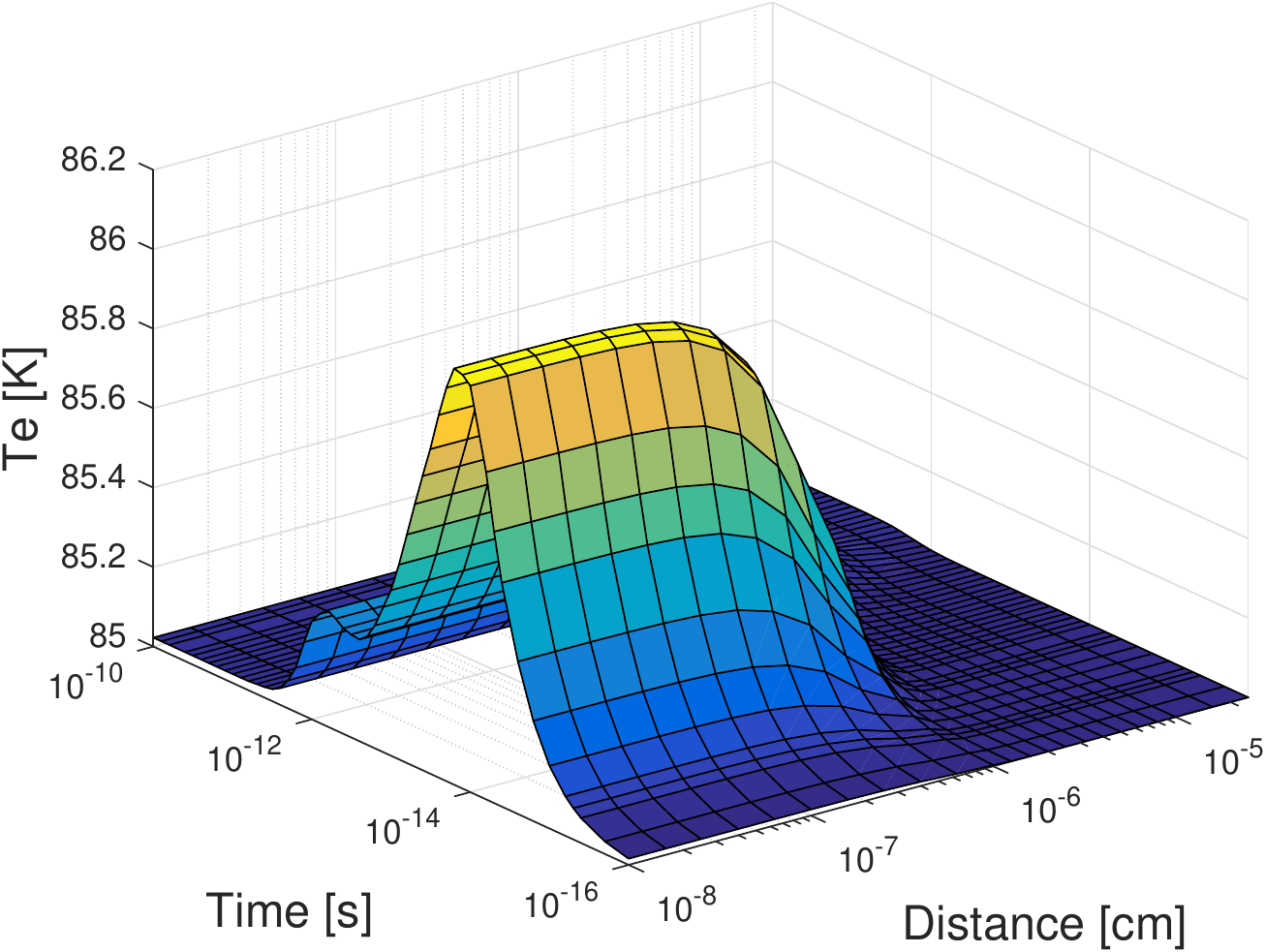}
  \caption{}
  \label{Fig6b}
\end{subfigure}
\caption{Time and space variation of the temperature of the molecular (a) and electronic system (b) of LAr produced by an Ar selfrecoil of 50 keV kinetic energy, resulting from the interaction of a $\mu$BH of velocity 250 km/s, mass 10$^{-5}$ g and charge number 30 with an Ar nucleus, at maximum energy transfer.}
\end{figure}

The thermal spike produced by an Ar self-recoil of kinetic energy 300 keV, produced by the collision with the $\mu$BH of velocity 250 km/s and mass $10^{-5}$ g at maximum transferred energy is represented in Fig. \ref{Fig6a} and \ref{Fig6b}.

The main difference with respect to the thermal spike produced by the $\mu$BH consists in the fact that for the Ar selfrecoil the electronic energy loss is higher, and it produces the most important peak in the electronic temperature at times characteristic to the electronic sub-system, while the peak corresponding to the transfer for the molecular sub-system is comparatively lower. 

The energy transferred from the molecular to the electronic subsystem on unit range has defined \cite{Lazanu:2015qpa} as
\begin{equation}
\left(\frac{dE}{dx}\right)_{ex}=\int_{0}^{\infty}{dt\int_{0}^{\infty}{2\pi rg\left(T_{mol}-T_e\right)\ dr}} .
\end{equation}

\begin{figure}[h]
\includegraphics[scale = 1]{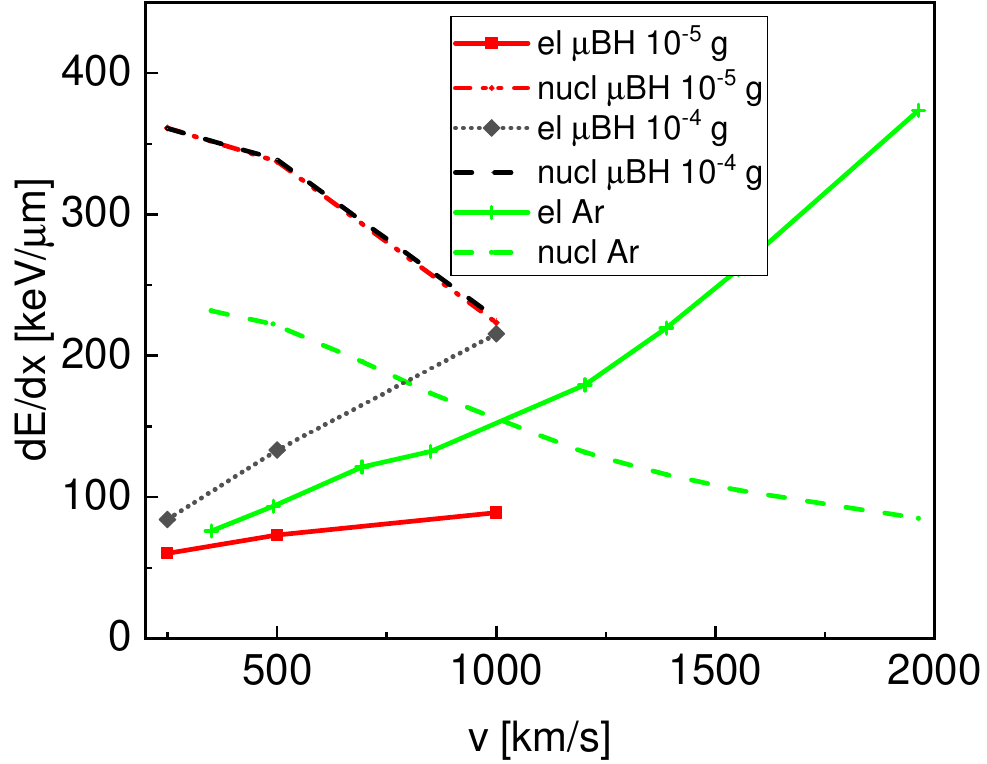}
\centering
\caption{Electronic (lines) and nuclear (dashed lines) energy loss for the $\mu$BH and Ar selfrecoils in LAr after the exchange of energy from nuclear to electronic subsystem.}
\label{Fig7}
\end{figure}

In Fig. \ref{Fig7}, the energy loss in LAr of $\mu$BHs having the masses of $10^{-5}$ and $10^{-4}$ g, the charge number 30 and the velocity $250 - 1000$ km/s is represented as a function of  velocity, taking into account the energy transferred by the molecular system to the electronic one during transient processes. On the same graph, the energy loss (electronic and nuclear) of the Ar selfrecoils produced by the interaction of the $\mu$BH in the target is also represented, also following the exchange of energy. One can see that the increase of the mass of the $\mu$BH conduces to the increase of the electronic energy loss, while the nuclear energy loss is practically unchanged. 

\begin{figure}[h]
\includegraphics[scale = 1]{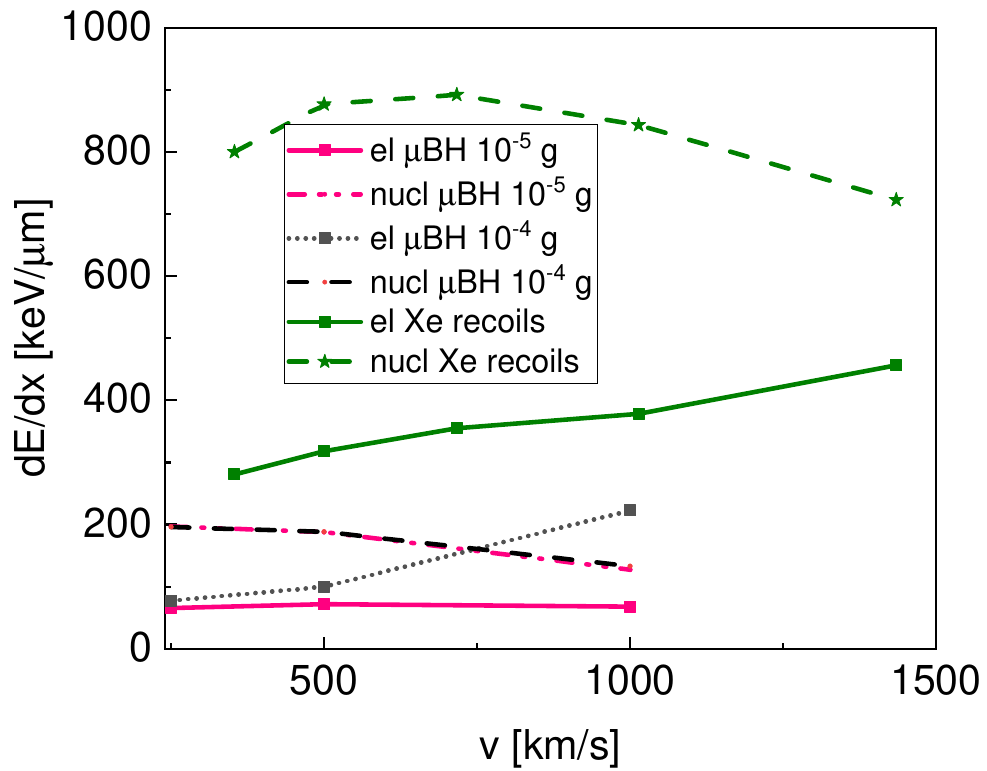}
\centering
\caption{Energy loss for the $\mu$BH (masses of $10^{-5}$ and $10^{-4}$ g, charge number 30 and velocity 250 – 1000 km/s) in LXe and for its selfrecoils after the exchange of energy between molecular and electronic subsystems during transient processes.}
\label{Fig8}
\end{figure}

For comparison, the same calculations have been performed for $\mu$BHs having the same characteristics in LXe. The energy loss as a function of velocity is represented in Fig. \ref{Fig8}, together with the energy loss of Xe selfrecoils produced by the $\mu$BH collisions in the target.

In all cases examined here, for the direct interactions of the $\mu$BH as well as for the selfrecoils it produces in LAr and LXe, the direction of the energy transfer is the same, \textit{i.e.} the electronic subsystem receives energy in transient processes in the investigated range of velocities, this being beneficial for the detection of the $\mu$BH. These transferred energies significantly modify the initial Lindhard partition of the energy loss.
For both LAr and LXe, the electronic energy loss is higher for the selfrecoils than for the $\mu$BH, and the electronic energy loss increases with the increase of the mass of the $\mu$BH.

Although the energy lost by the $\mu$BH during its passage through LAr is negligible with respect to its energy, the energy deposited in the electronic subsystem produces ionizations and excitations that are at the base of its detection.

Thus, for a $\mu$BH of velocity 250 km/s and mass 10$^{-5}$ g, our calculations revealed that, following the energy transfer from the molecular subsystem, (\textit{dE}/\textit{dx})$_{el}$ is 60 keV/$\mu$m. The average energy transferred to an Ar recoil in a head-in collision is 25 keV, which eventually losses 76 keV/$\mu$m as excitations and ionizations, by considering also the transfer from the molecular subsystem. Thus, 6.97 $\times$ 10$^3$ photons/$\mu$m are estimated and this maximal component will contribute to scintillations.

Consequently, in the analysis of tracks in LAr and LXe detectors, the trajectories of these $\mu$BHs will appear as crossing the whole active medium, in any direction, producing uniform ionization and scintillation for the whole path.

\section{Summary}
We investigated the possibility of direct detection of hypothetical objects formed just after the Big Bang, and which survived until now. We considered the case of black holes with masses around the Planck scale (10$^{-5} - 10^{-4}$ g) which ensure classical gravitational treatment of the objects. 
The $\mu$BHs are viable candidates in explaining the nature of dark matter. Originally proposed by Stephen Hawking in 1971, in recent years these objects became an important option especially when other highly sought-after candidates have not been discovered.
In this work, based on the idea of their existence, we considered that such relics carry electric charges up to 30 in units of electron charge. We discussed some arguments in favour of the supposition that they do not radiate and thus they have survived until today. 
The results obtained for their electronic and nuclear stopping powers in LAr, calculated in the frame of the Lindhard and Ziegler formalisms, including the transient processes treated in the frame of the thermal spike model, considering the transfer of energy between electronic and molecular subsystems, open the possibility of their direct detection in future generation of huge LAr detectors. These results are applicable also to LXe detectors. The main uncertainty for more realistic predictions is the flux of these objects. Our estimations show that the discrimination between the signals (ionization and scintillations) produced by these $\mu$BHs and other heavy ions or other particles is possible. There is a clear distinction between expected signals from charged $\mu$BH objects and WIMPs for example.

\acknowledgments

IL and MP acknowledge the support from the Romanian Programme PNCDI III, CERN-RO, under Contract 2/2020, and SL of NIMP Core Program No. 21N/2019.

\bibliography{Bibliography}{}

\end{document}